# Coherent, super resolved radar beamforming using self-supervised learning

Itai Orr[1,2*], Moshik Cohen[2], Harel Damari[2], Meir Halachmi[2], Zeev Zalevsky[1]

[1]Faculty of Engineering and the Institute for Nanotechnology and Advanced Materials, Bar Ilan University, Ramat-Gan, Israel
[2]Wisense Technologies Ltd., Tel Aviv, Israel

**Abstract:**
High resolution automotive radar sensors are required in order to meet the high bar of autonomous vehicles needs and regulations. However, current radar systems are limited in their angular resolution causing a technological gap. An industry and academic trend to improve angular resolution by increasing the number of physical channels, also increases system complexity, requires sensitive calibration processes, lowers robustness to hardware malfunctions and drives higher costs. We offer an alternative approach, named **R**adar signal **R**econstruction using **S**elf **S**upervision (R2-S2), which significantly improves the angular resolution of a given radar array without increasing the number of physical channels. R2-S2 is a family of algorithms which use a Deep Neural Network (DNN) with complex range-Doppler radar data as input and trained in a self-supervised method using a loss function which operates in multiple data representation spaces. Improvement of 4x in angular resolution was demonstrated using a real-world dataset collected in urban and highway environments during clear and rainy weather conditions.

## INTRODUCTION

Autonomous vehicles attracted great attention in recent years due to their tremendous impact on the economy and society (*1*) as well as their potential to save lives (*2*). The evolution from current driver assistance systems into fully autonomous vehicles requires several, functionally independent sensing modalities for real time sensing and perception (*3*). The requirement for sensing redundancy (*4*) spurred research toward more advanced camera and LiDAR based solutions. However, these sensing modalities suffer from inherent sensitivity to harsh weather and limited effective range, due to the electromagnetic spectrum they utilize, usually $400 - 800 nm$ for cameras and $850 - 950 nm$ or $1.45 - 1.55 \mu m$ for LiDARs.

In contrast, automotive radar usually utilizes a frequency spectrum of $76 - 81 GHz$, which offers robustness to weather conditions as well as longer effective range. However, utilization of radar for autonomous driving is hindered in part due to the relatively limited angular resolution currently provided by available commercial platforms.

The angular resolution of a radar translates to the ability to distinguish and separate between targets and is proportional to the antenna diameter. In automotive scenarios, where the environment is usually rich with objects and targets (i.e. cluttered environment), angular resolution is critical. For example, two cars driving in adjacent lanes might be miss-detected as a single object by a limited angular resolution radar.

In a radar array, the individual antenna elements are usually positioned about $\lambda/2$ apart from each other, with $\lambda$ representing the central wavelength in free-space; therefore increasing the number of antenna elements should enlarge the dimensions of the physical aperture. Following this principal, an industry and academic trend has emerged to enlarge the aperture by increasing the number of physical transmitting and receiving channels. The drawbacks of this approach are complex system architecture prone to hardware failure, requirement for sensitive calibration process and high costs which hinder the adaptation of such systems in commercial applications.

An additional important factor affecting a radar's angular resolution is the algorithm used for beamforming. Fast Fourier Transform (FFT) performed on the angular dimensions of a radar array is considered a conventional beamformer and sets the Fourier resolution of a radar. Super-resolution (SR) methods which aim to achieve sub-Fourier resolution, include Estimation of Signal Parameters via Rotation Invariance Techniques (5) (ESPRIT) or the popular Multiple Signal Classification (6) (MUSIC). MUSIC's main disadvantages are a requirement of prior information on the number of targets, assumption on coexistent targets to be uncorrelated and high computation costs. These limitations make its use in real-world automotive radar applications more challenging. In addition, most current SR methods usually require using several snapshots (i.e. frames) in order to improve the estimation of the spatial covariance matrix. This requirement is problematic in safety critical, automotive applications since each added snapshot increases the response time of the system.

Recently, deep learning has begun to make an impact on traditional radar signal processing, perception and system design. Radar data was used with DNN for high resolution road segmentation (7), road user classification (8), multi-class object classification (9), road user detection (10), vehicle detection (11), lane detection (12) and semantic segmentation (13, 14). Apart from perception tasks, DNNs have proven useful for cognitive antenna design in phased array radar (15) and enhanced radar imaging (16).

Another class of algorithms in radar signal processing is Compressed Sensing (CS), which usually exploit sparseness in a scene to reconstruct one or more dimensions of a radar data tensor (i.e. range-Doppler-azimuth-elevation). However, this is a property that does not often occur in cluttered, urban driving scenarios. Further details on the lack of sparsity in the dataset used in this work are provided in the supplementary material and shown in Fig. S1. Complex Block Sparse Bayesian Learning was demonstrated for radar signal reconstruction (17). A spatial CS framework (18) was developed and evaluated by numerical simulations for 5 targets with constant Signal to Noise Ratio (SNR). The authors assumed the number of targets is known, noise level is available and forewent from estimation of measurements in the range and Doppler dimensions. Iterative method with adaptive thresholding (19) was used for a sparse Multiple In Multiple Out (MIMO) radar array where the authors (20) conducted examinations of a corner reflector in an anechoic chamber and a single parked vehicle at a range of 4m. Examination of CS for MIMO radar (21) concluded that these techniques remain valid when there are under 106 scatter points in a scene. However, in typical urban scenes which may contain many more scatter points, these methods require using a high SNR threshold in order to minimize the number of scatterers.

Research towards utilizing DNNs to improve radar angular resolution is in its early stages. Radar data in range-Doppler representation was used with a Generative Adversarial

Network (GAN) architecture to demonstrate SR in two specific cases (*22*): pedestrian micro-Doppler signature by collecting data of people walking on a treadmill and a staircase which achieved angular SR with a factor of $2x$. The authors (*22*) point out the difficulty of assembling a large manually labelled dataset in real-world scenarios for the general case of numerous types of objects, classes, materials and shapes.

Instead of real-world data, synthetic data was used for training with a single radar snapshot as input (*23*). However, using synthetic data for training before deployment in a real-world environment usually results in relatively lower performance caused by modelling and numerical errors in the simulation used to create the synthetic data. This is also referred to as the sim-to-real adaptation challenge.

Multiple snapshots of a spatial covariance matrix were used with a Convolutional Neural Network (CNN) and a 1D antenna array with simulated data for Direction of Arrival (DOA) estimation and SR (*24*). A single snapshot of a spatial covariance matrix was used with a fully connected model for DOA estimation and super resolution of a 2D antenna array with simulated data (*25*) and a 1D antenna array with both simulation and real-world data where the targets were corner reflectors (*26*). Two snapshots were used with an anechoic chamber setup to generate a dataset which was used with a fully connected model for DOA estimation (*27*).

Although shown only for simulated data or controlled scenarios with very few targets, these works show the potential DNN have for super-resolving radar arrays in real-world environments which usually contain many targets and reflections. We hypothesize that previous methods for DNN-based radar SR failed or did not try to generalize to uncontrolled, real-world environments mainly due to a lack of a suitable training methodology.

Self-supervised learning is a young research area and is considered a part of unsupervised training, where one part of a data is used to predict a different part of the same data. The strength and disruptive potential of this training methodology lies in the fact that in many applications, data is in abundance, however labeling the data, which is essential for supervised training, is a time consuming and expensive process. Furthermore, in some applications such as image denoising (*28*), manual labeling is not a viable solution. Self-supervised techniques showed promising early results for semantic image segmentation (*29–31*), temporal cycle-consistency to learn temporal alignment between videos (*32*), dense shape correspondence for 3D objects (*33*) and feature representation for visual tasks (*34–36*).

The field of image SR has also utilized self-supervision to create State-Of-The-Art (SOTA) result (*37–40*). At its fundamentals, self-supervision for image SR uses a high-resolution image which is down-sampled to create a low-resolution image. A DNN is then trained using the low-resolution image as input and the high-resolution image as label.

Apart from computer vision, research into signal processing has also begun using self-supervised learning with other forms of data. In audio data, it was used for speech enhancement (*41*), pitch estimation (*42, 43*), source separation (*44*) and feature representation (*45–47*). In electroencephalography data, self-supervision was used for representation learning (*48*) and in electrocardiogram data, self-supervised learning was utilized for emotion recognition (*49*).

This work proposes to leverage self-supervised learning to super-resolve a radar array. More specifically, R2-S2 uses an auto-encoder trained in a self-supervised method with a diluted radar array and used to reconstruct the amplitude and phase of missing receiving channels, where the dilution of the radar array was designed to limit the resolution of the input array. To enforce coherence during the reconstruction process, a loss function which operate on multiple data representation spaces was utilized.

In contrast to several radar-based CS and SR methods, R2-S2 does not require sparsity in the range-Doppler-azimuth dimensions, it can be used in highly cluttered environments such as crowded urban streets with numerous objects and targets present in the radar Field of View (FOV) and does not require prior knowledge on the number of targets in a scene. Validation was performed on a real-world dataset collected using a vehicle mounting a radar unit and driven in urban and highway environments in both clear and rainy weather conditions.

## RESULTS

### Data

In this work we target the general application of radar SR. To demonstrate our approach, a dataset was collected in uncontrolled urban and highway environments in both clear and rainy weather conditions, using a vehicle mounting a temporally synced camera and radar with their field of view overlapped. The dataset was split into 85,671 frames for training and 6,632 frames for validation. The validation dataset was separated from the training dataset by collecting data during different dates and locations in order to avoid the appearance of similar frames in both datasets, which could have occurred in the case of simple random split. Samples from the training dataset are shown in Fig. 1.

We used a Frequency Modulated Continuous Wave (FMCW) MIMO radar with a 79GHz carrier frequency. A FMCW radar transmits a linear chirp signal whose frequency increases linearly with time. When combined with means of signal processing (mainly FFT), it is possible to extract useful information from the raw signal such as, range, velocity and DOA (*50*).

MIMO radar is comprised of multiple transmitters (Tx) and receivers (Rx) antennas. Each transmitter can transmit a waveform independently of the other transmitting antennas while each of the receiving antennas can also receive these signals independently. By processing measurements from different transmitting and receiving antennas, one can create a virtual aperture whose size is larger than the physical aperture. i.e. an antenna array comprised of $NTx$ transmitters and an array of $NRx$ receivers, results in a virtual array of $NTx{\times}NRx$ channels. This increase in aperture size, translates to improved performance such as: spatial resolution, resistance to interference and probability of detection of the targets (*51*). In this work a collocated MIMO radar was used, however, the proposed method can be applied to a non-collocated MIMO radar as well. In addition, although we focus on MIMO radar in this work, a similar approach can be applied with other multi-channel radars.

### Data preprocessing

A radar signal in its raw form, contains a variety of information originating from different physical phenomena in the environment such as targets reflections and electromagnetic wave propagation through the atmosphere. In addition, hardware related effects originating from components such as the signal generator, receive/transmit chains and antenna elements also greatly impact data fidelity. The combination of numerous, simultaneous, sometimes non-linear and often coupled mechanisms which affect a radar signal, also make it mathematical modeling very difficult (*52, 53*). To differentiate between different interactions, FFT has long become a staple for radar signal processing. More specifically, FFT is used to transform a signal from its raw measurement form to different representation spaces, such as range-Doppler for example. In this work, the radar used a Uniform Linear Array (ULA) antenna array configuration with 16 virtual channels, providing the ability to process both amplitude and phase information in 3 dimensions: range, Doppler and azimuth. The waveform used was configured to 48 sweeps and 256 samples with maximum detection range of 64m and maximal relative velocity of 5.8m/s. While the FOV was configured to $100^o$ horizontal.

The input to the model was created by applying a window function and FFT on both sweeps and samples dimensions to generate a complex data tensor with the dimensions of virtual channel, range and Doppler. More specifically, the original signal $x_{raw}$ has the dimensions of (virtual channel, samples, sweeps) and first goes through the range processing described in eq. 1 which includes windowing and real to complex FFT on the sample dimension.

$$x_{range} = \mathcal{F}_{sample}\big(W_{sample}(x_{raw})\big) \tag{1}$$

Where $\mathcal{F}$ is FFT and $W$ is a window function. The transformed signal $x_{range}$ has the dimensions of (virtual channel, range, sweeps) and goes through Doppler processing which includes windowing and complex to complex FFT on the sweeps dimension, as described in eq. 2.

$$x_{range-Doppler} = \mathcal{F}_{sweep}\big(W_{sweep}(x_{range})\big) \tag{2}$$

The resulting signal $x_{range-Doppler}$ has the dimensions of (virtual channel, range, Doppler) and is then used as input to a DNN. The suggested method holds several important characteristics in regard to data preprocessing which contribute to its generality and robustness while addressing the shortcoming of previous approaches to radar SR. Mainly, there is no requirement for specific filtering, there are no assumptions on the sparsity of the data, there is no minimum SNR threshold, no calibration is required, there is no maximum number of scatter points and there are no requirements of prior information on the scene. In addition, in order to remove the requirement for complex radar signal modelling, R2-S2 was designed as an end-to-end approach, forcing a DNN to implicitly learn a signal model as part the reconstruction process. Meaning, there is no need to provide an accurate and detailed mathematical description of the signal.

## Experiments
The model was given as input a diluted 1D sub-array of complex (both amplitude and phase) range-Doppler maps while the remainder array was used as label. Meaning, the training is performed in self-supervised manner. As there are numerous possible permutations for the choice between input and label receiving channels, experiments were performed using an example configuration described in the 'Materials and Methods' section and shown in Fig. 2 where a virtual array of 16 channels was split to 4 receiving channels used as an input array while the remainder 12 receiving channels are used as label. Meaning, the combined array has 4x improved resolution than the input array.

Sample results from the validation dataset are provided in Fig. 3 showing representative scenarios from urban and highway environments in both clear and rainy weather conditions. Also provided in Fig. 3, cartesian view comparison between the label and predicted beamformers which were obtained by performing FFT on the channel dimension of the original array and the predicted array as shown in Fig. 2. These results demonstrate the use of R2-S2 to super-resolve a limited angular resolution RADAR array thereby achieving 4x improved resolution in scenarios representing various combinations of dynamic and static objects, including vehicles, vegetation, sidewalks, poles and structures.

Further validation was performed using two evaluation metrics: L1 and PSNR. Both were averaged over the validation dataset. Lower L1 error corresponds to improved reconstruction and was calculated by eq. 3:

$$\text{L1} = \frac{1}{N_i N_j} \sum_{i,j} \frac{\left| y_{i,j}^{pred} - y_{i,j}^{label} \right|}{\left| y_{i,j}^{label} \right|} \tag{3}$$

Where $L1$ is the reconstruction metric. In the range-Doppler representation space both metrics were calculated for each receiving channel separately while in the beamformer representation space (i.e. rang-Doppler-azimuth) the metrics were calculated globally to focus on coherence.

Since R2-S2 deals with coherent reconstruction of an array's response, the important metrics are associated with the beamformer representation space and more specifically, the combination of low L1 and high PSNR which correlate to coherent beamforming.

Table 1 displays an ablation study performed on the loss function described in the 'Materials and Methods' section and was averaged over the validation dataset. The results show that the best performances (in bold) are achieved by using all parts of the loss function, suggesting improved coherence is attained by adding the beamformer constraints to the optimization process.

The critical importance of superior angular resolution for automotive radars can be further understood by examining common everyday driving scenarios as demonstrated in Fig. 4. These examples demonstrate how limited-resolution radars (i.e. the input radar array used) can falsely detect objects in front of the vehicle even though the road ahead is clear. In addition, adjacent objects can also be falsely detected as a single object. These highly undesired phenomena can be resolved by using our method to increase the angular resolution of the radar array.

To further support the general applications of R2-S2, experiments were performed with a different permutation of input and label receiving channels. This configuration, which was called 'sparse array configuration' is displayed in Fig. 5 where R2-S2 is used to interpolate receiving channels between sparsely spaced input receiving channels. Sample results from the validation dataset are provided in Fig. 6, where 4 uniformly spaced receiving channels are used as input and 12 receiving channels are used as label. In this configuration, the resolution of the input and label arrays are similar (they share aperture size), however due to the large spacing between receiving antenna elements in the input array, the input beamformer suffers from high grating lobes which severely degrade performance. By applying R2-S2 we were able to coherently reconstruct the missing receiving channels and match the performance of the label array.

Additional validation of the sparse array configuration was performed on the validation dataset and compared to bi-cubic interpolation. The results provided in Table 2 show that bi-cubic interpolation does not enforce coherence during the reconstruction process, as evident by the high L1 score in the beamformer representation space. In contrast, R2-S2 is able to reconstruct the array correctly and coherently.

Since R2-S2 uses signal reconstruction to improve resolution, it can also be used for mitigation of hardware failure. More specifically, in cases where one or more receiving channels are randomly corrupted, the suggested method can be used to replace them with artificial receiving channels. This configuration, which was called 'random missing channels configuration' is displayed in Fig. S2.

To demonstrate this approach, experiments were performed where a DNN is used to estimate random missing receiving channels. Since the number and position of the missing receiving channels can vary and is not known in advance, the DNN first needs to determine if each receiving channel is corrupt and then coherently reconstruct it based on the remaining receiving channels. To assess the performance of this configuration we first conduct a quantitative comparison to bi-cubic interpolation with a single randomly missing receiving channel. The results are provided in Table S1 and were performed using the validation dataset, where R2-S2 outperforms bi-cubic interpolation as evident in lower L1 and high PSNR in the beamformer representation space. Note that bi-cubic interpolation cannot estimate receiving channels at the edge of an array whereas our method is able to extrapolate as well as interpolate.

A generalized version of this configuration utilizing a single DNN trained to predict a random number of randomly positioned receiving channels was also assessed. A quantitative comparison of a DNN trained with up to 8 random missing channels and validated over the validation dataset is shown in Fig. S3. As expected, we observe a decrease in performance as reflected in L1 and PSNR metrics as the number of random missing channels increases. Sample results from the validation dataset are provided in Fig. S4 showing a detailed analysis of R2-S2 for the random missing channels configuration where we observe that R2-S2 can reconstruct the range-Doppler maps and create a coherent array. By combining this analysis with a specific performance criterion, it is possible to set a maximum number of missing channels for which this configuration may be used for in a real-world application.

## DISCUSSION

Limited angular resolution is one of the main limiting factors in automotive radar applications. An industry trend to improve angular resolution by increasing the number of physical receiving channels also increases system complexity, creates cumbersome calibration processes, adds sensitivity to hardware failure, decreases power efficiency and drives higher cost. An alternative approach is to use SR algorithms. However, unless very carefully designed and implemented, this can also introduce sensitivity to calibration, increase latency, add limitations on the number of targets and in some cases a requirement for prior knowledge on the environment.

To address these limitations, R2-S2 was designed with a single snapshot as input which is an important property in automotive applications where reaction time is critical. Furthermore, the dataset was collected in uncontrolled urban and highway environments during both clear and rainy weather conditions and was not focused on a specific class of objects. The preprocessing stage did not contain special filtering nor requires any calibration process, there is no requirement for prior knowledge on the number of targets in a scene and no minimum SNR threshold. In addition, the run-time is invariant to the number of detections in a frame. Meaning, a highly cluttered scene will not cause a bottleneck in processing time which is an important characteristic in real-time applications.

The proposed approach can replace or used in addition to existing SR methods and uses self-supervised learning to train a DNN to predict artificial receiving channels in range-Doppler representation outside of an array's aperture. The combined, original and artificial receiving channels create a larger aperture, if coherence is maintained, the improvements of the larger array are improved angular resolution and higher SNR.

To enforce coherence, additional constraints were introduced during the training process. These constraints were in the form of additional loss terms operating in the beamformer representation space. Training was performed using both representation spaces (i.e. range-Doppler and beamformer representations) simultaneously.

In this work, FFT was chosen as a beamformer. However, alternative beamformers can also be used. For example, the constraints introduced in the loss function as $\mathcal{L}_{bf}$ can be created by applying SR algorithm such as MUSIC. By combining the suggested approach with other SR methods, it may be possible to achieve higher improvement factors than previously achieved.

Experiments were performed with a configuration of 4 input receiving channels and 12 label receiving channels which achieved a 4x improved angular resolution factor. However, additional permutations are also possible, for example, 8 input receiving channels and 8 labels receiving channels would have created a 2x improved angular resolution factor. Furthermore, given a larger original radar array, the suggested method can potentially achieve larger improvement factors. For example, an array with 64 receiving channels can be split into 8 input receiving channels and 56 label receiving channels which results in 8x improved resolution factor.

An important observation is shown in Fig. 4c, which demonstrates a case where the radar is stationary as evident by the Doppler plot centered around $v = 0\ m/s$. In this case, similar qualitative results arise in comparison to cases where the radar was moving, which suggests that in contrast to previous methods (22), a DNN trained with R2-S2 does not rely exclusively on the Doppler and micro-Doppler effects during the reconstruction process.

R2-S2 can also be used with different a type of configuration, which was called 'sparse array' and simulates a sparse radar array. Meaning, the distance between each virtual antenna element is larger than $\lambda/2$, which is optimal in terms of grating lobes and spatial ambiguity. This allows the array to have a larger aperture size, thus improving its angular resolution. However, this enlarged element distance causes degraded performance in the element pattern of the array as observed in Fig. 6. Where we demonstrate that when only using the sparse input receiving channels for beamforming, there is a significant reduction in SNR compared to using the entire array. However, by utilizing R2-S2, coherent artificial receiving channels are predicted to fill in the gaps, which make it possible to have a larger aperture and still maintain high performance, matching those of the full array.

Additional application of R2-S2 refers to its use for mitigation in cases of corrupt receiving channels. This configuration trains a DNN to predict one or more random receiving channels from the remaining functional receiving channels. During inference, the missing receiving channels can be any receiving channel in the array, without the need to change configuration or 'notify' the model which receiving channel is missing. Meaning, the model identifies which receiving channel is missing and predicts the appropriated artificial receiving channel.

This work offers an alternative approach to conventional radar beamforming and super resolution which challenges an industry and academic trend towards increasing the number of physical channels in radar arrays in order to achieve improved angular resolution. The suggested method, termed R2-S2, uses a DNN trained in a self-supervised method with a diluted antenna array to super-resolve a radar by coherently predicting the amplitude and phase of receiving channels outside of the physical or virtual aperture using a novel loss function in multiple data representation spaces. The results demonstrated robust, real time performance and an improvement factor of 4x in cluttered scenarios by using a real-world dataset collected in urban and highway environments during clear and rainy weather conditions. In addition, R2-S2 can also be used for mitigation of hardware failure which can further increase the reliability of automotive radars.

This work suggests that learning based methods can be combined or replace traditional methods for radar super-resolution in real-world applications. We hope our method will assist to bridge the technological gap in radar angular resolution and enable radar centric autonomous driving. In a broader sense, this work demonstrates how self-supervised learning can be used for radar signal processing which we hope will inspire more research in this direction.

## MATERIALS AND METHODS

### Training methodology

A fundamental concept of self-supervised learning involves manipulating, augmenting or masking parts of an input data and then predicting the original data, part of the original data or which manipulation was performed. In this work we propose to use self-supervision to predict radar data and treat the SR problem as a signal reconstruction problem while combining it with traditional beamformers. Meaning, our method can work in combination with other SR methods.

In order to improve a radar array's angular resolution, R2-S2 uses a DNN to predict received data outside of the physical or virtual array aperture. The combination of the original receiving channels and the predicted receiving channels create an artificial radar array with a larger aperture and thus improved angular resolution. In this work, the term artificial receiving channels is used for the predicted receiving channels in order to differentiate them from virtual receiving channels created by a MIMO process.

We propose to expand a virtual MIMO array and create an artificial array comprised of virtual receiving channels from the MIMO array and artificial receiving channels from the DNN's prediction. Together, all channels can then be used with a beamformer. In this work, FFT was used as beamformer, however R2-S2 can also be applied with other beamformers, for example MUSIC or ESPRIT. The coherence of the predicted artificial channels requires special attention. Otherwise, the resulting beamforming will not achieve SR. This task is especially difficult since it requires a DNN to extrapolate coherent data for receiving channels positioned far from the original input receiving channels.

Our method allows for flexibility in the partitioning between input and label receiving channels. For example, in a ULA with 16 channels, a partitioning of 8 input receiving channels and 8 label receiving channels will result in a 2x angular resolution improvement factor. Another example for the receiving channels partitioning is displayed in Fig. 2, where a ULA with 16 receiving channels is considered. The central 4 receiving channels are used as input to a DNN while the remainder 12 receiving channels are used as label, as seen in Fig. 2a. During inference mode, the original input receiving channels are used twice, first as an input to a DNN from which 12 receiving channels are predicted. Second, they are used together with the predicted receiving channels to create an artificial array which has a total of 16 receiving channels, thus 4x improved resolution from the original 4 receiving channels input array, as seen in Fig. 2b.

**Model**
The model used in all experiments was adapted from (7) and based on the encoder-decoder Unet (54) model combined with position embedding and self-attention (55) layers working on the channel dimension to encourage learned cross channel correlations. Additional layers used were average pooling, leaky-Relu activation and instance normalization. All convolution and transpose convolution used a 3x3 kernel. The proposed model has about 1.4M parameters and achieves 15ms inference time on 2080Ti GPU, which makes the suggested approach attractive for embedded, real time applications.

**Loss Function**
In order to coherently reconstruct a radar array's response, a loss function was constructed which operates in two data representation spaces simultaneously. As a general partitioning, the first loss representation space was range-Doppler ($\mathcal{L}_{rd}$) and was used to reconstruct the amplitude. The second loss representation space, termed 'Beamformer' ($\mathcal{L}_{bf}$), was achieved by applying FFT on the channel dimension and was used mainly to reconstruct the phase while enforcing coherence throughout the array.

The loss term is a sum of range-Doppler based and beamformer based losses: $\mathcal{L} = \mathcal{L}_{rd} + \mathcal{L}_{bf}$. The resulting multi-objective loss function combines two different physical representation, therefore addition of such loss terms should be done carefully. During experimentation, normalization of each loss term was examined, however, no significant performance improvements were observed.

Both loss terms are composed of three components: reconstruction loss, energy conservation and total variation. In the range-Doppler representation space, the loss function is displayed in eq. 4:

$$\mathcal{L}_{rd} = \lambda_{rd_{rec}} \mathcal{L}_{rd_{rec}} + \lambda_{rd_{energy}} \mathcal{L}_{rd_{energy}} + \lambda_{rd_{tv}} \mathcal{L}_{rd_{tv}} \tag{4}$$

Where ($\lambda_{rd_{rec}}, \lambda_{rd_{energy}}, \lambda_{rd_{tv}}$) are hyperparameters for the reconstruction, energy and total variation losses respectively. $\mathcal{L}_{rd_{rec}}$ is the L2 reconstruction loss, shown in eq. 5:

$$\mathcal{L}_{rd_{rec}} = \frac{1}{N_i N_j} \sum_{i,j} \left(y_{i,j}^{pred} - y_{i,j}^{label}\right)^2 \tag{5}$$

$N_i$ is the number of samples, $N_j$ is the number of receiving channels, $y_{i,j}^{pred}$ is the DNN prediction for sample $i$ of a receiving channel $j$ in range-Doppler representation and $y_{i,j}^{label}$ is the associated label. $\mathcal{L}_{rd_{energy}}$ is a smooth L1 energy conservation loss, shown in eq. 6,7:

$$\mathcal{L}_{rd_{energy}} = \frac{1}{N_i N_j} \sum_{i,j} z_{i,j} \tag{6}$$

$$z_{i,j} = \begin{cases} 0.5 \cdot \left(\left|y_{i,j}^{pred}\right| - \left|y_{i,j}^{label}\right|\right)^2 & if \ \left|\left|y_{i,j}^{pred}\right| - \left|y_{i,j}^{label}\right|\right| < 0.5 \\ \left|\left|y_{i,j}^{pred}\right| - \left|y_{i,j}^{label}\right|\right| - 0.5 & otherwise \end{cases} \tag{7}$$

$\left|y_{i,j}^{pred}\right|$ is the amplitude of the DNN's prediction. Displayed in eq. 8 and 9, $\mathcal{L}_{rd_{tv}}$ is the total variation loss calculated over the range and Doppler dimensions:

$$\mathcal{L}_{rd_{tv}} = \frac{1}{N_i N_j} \sum_{i,j} tv_{i,j} \tag{8}$$

$$tv_{i,j} = \frac{1}{N_k N_l} \sum_{k,l} \left|\left|y_{i,j}^{pred}(k,l)\right| - \left|y_{i,j}^{pred}(k-1,l-1)\right|\right| \tag{9}$$

Where $(N_k, N_l)$ are the number of range and Doppler bins respectively. All three loss terms were calculated per receiving channel separately to enforce tighter constraints and facilitate better reconstruction results.

The second loss term $\mathcal{L}_{bf}$ , which operates in the beamformer representation space was calculated with similar expressions for the reconstruction, energy conservation and total variation losses. Key differences were made in order to encourage correct phase reconstruction. Here, the reconstruction loss $\mathcal{L}_{bf_{rec}}$ is calculated globally, in order to enforce coherence between the different channels as shown in eq. 10:

$$\mathcal{L}_{bf_{rec}} = \frac{1}{N_i} \sum_i \left(y_i^{pred} - y_i^{label}\right)^2 \tag{10}$$

In addition, energy conservation loss $\mathcal{L}_{bf_{energy}}$ is calculated per azimuth bin, as shown in eq. 11:

$$\mathcal{L}_{bf_{energy}} = \frac{1}{N_i N_m} \sum_{i,m} z_{i,m} \tag{11}$$

Where $N_m$ is the number of azimuth bins and $z_{i,m}$ is described in eq. 6. Total variation $\mathcal{L}_{bf_{tv}}$ was performed on the range and azimuth dimensions, as shown in eq. 12,13:

$$\mathcal{L}_{bf_{tv}} = \frac{1}{N_i N_l} \sum_{i,l} tv_{i,l} \tag{12}$$

$$tv_{i,l} = \frac{1}{N_k N_m} \sum_{k,m} \left|\left|y_{i,l}^{pred}(k,m)\right| - \left|y_{i,l}^{pred}(k-1,m-1)\right|\right| \tag{13}$$

**Implementation Details**

Training was implemented in Pytorch, optimizer used was Adam with $\beta_1 = 0.9, \beta_2 = 0.999$, batch size 16 and learning rate utilized cosine decay from $3.141 \cdot 10^{-4}$ to $3.141 \cdot 10^{-7}$. Training was continued until convergence and took about 30 epochs.

**Sparse array configuration**

Given a radar array, R2-S2 provides design flexibility in the partitioning between input and label receiving channels. As an additional example for this degree of freedom, an additional configuration is demonstrated in Fig. 5a where an array of 16 receiving channels is split into 4 input receiving channels spread uniformly across the original array and 12 label receiving channels. Inference mode for this configuration is shown in Fig. 5b, where the 4 input receiving channels are first used with a DNN to predict 12 coherent artificial receiving channels. Afterwards, both input and predicted receiving channels are arranged in their correct place in an array to allow for coherent beamforming.

This configuration is used to predict receiving channels in a MIMO virtual array based on neighboring channels. Meaning, a DNN is used to interpolate missing receiving channels in a MIMO virtual array. Performance improvement using this configuration can be achieved in two ways. First, given a specific performance metric, it is possible to decrease the number of receiving channels while still retaining high level of performance, thus saving cost and simplifying system architecture and design. Second, given a specific number of receiving channels, this configuration allows to increase the aperture size (thus improving the angular resolution) and retain coherent beamforming with high SNR and low sidelobes. This is achieved by rearranging the receiving channels and spreading them over a larger aperture size, which improves the angular resolution. However, simply increasing the distance between each receiving channel can decrease the array's performance significantly. For this end, a DNN is used to fill in the gaps with coherent artificial receiving channels and match the performance of a larger array.

**Random missing channels configuration**

In addition to SR, R2-S2 can also be used for other purposes. In scenarios where a receiving channel becomes corrupt or exhibits performance degradation during runtime operation, a DNN trained with our method can be used to replace the corrupt receiving channel with an artificial receiving channel. To accomplish this, R2-S2 is used to predict random missing receiving channels from the reminder active radar array. This task is especially difficult for a DNN since the receiving channels are randomly chosen and can also be located at the edges of the array, meaning the DNN needs to extrapolate as well as interpolate.

To create a DNN which is invariant to the position of a missing receiving channel a full MIMO virtual array is used as input and randomly chosen receiving channels are masked while the DNN is tasked to predict the missing receiving channels. The resulting trained DNN is invariant to the specific receiving channel missing and is able to reconstruct the data of each receiving channel individually without the need to train a separate model for each receiving channel. An illustration of the training methodology for this configuration is provided in Fig. S2.

**Figures:**

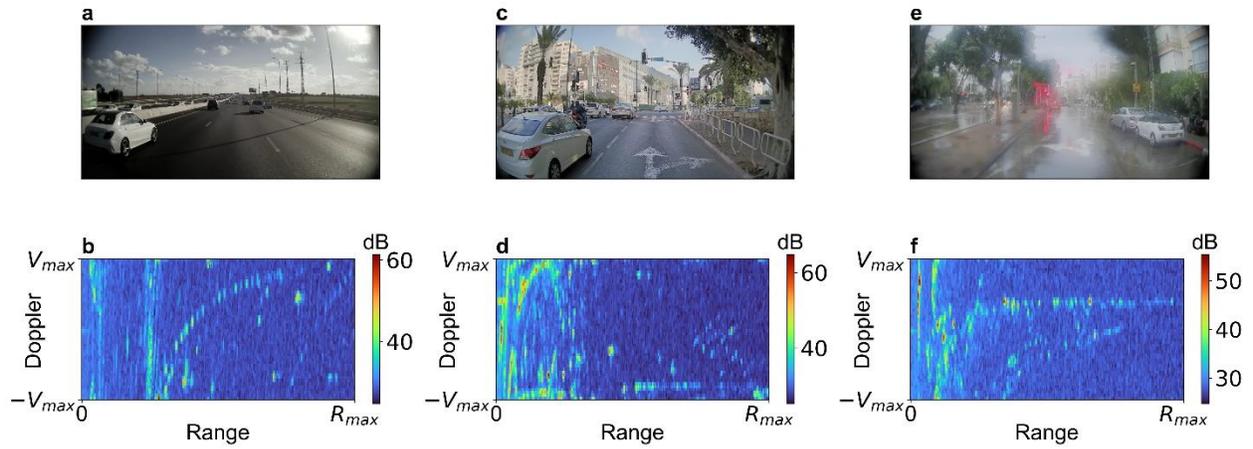

**Fig. 1. Sample frames from the training dataset a-f. (a,c,e)** camera image. **(b,d,f)** the respective range-Doppler map in dB. Blurry image in **e** was caused by rain droplets on the camera lens during data collection.

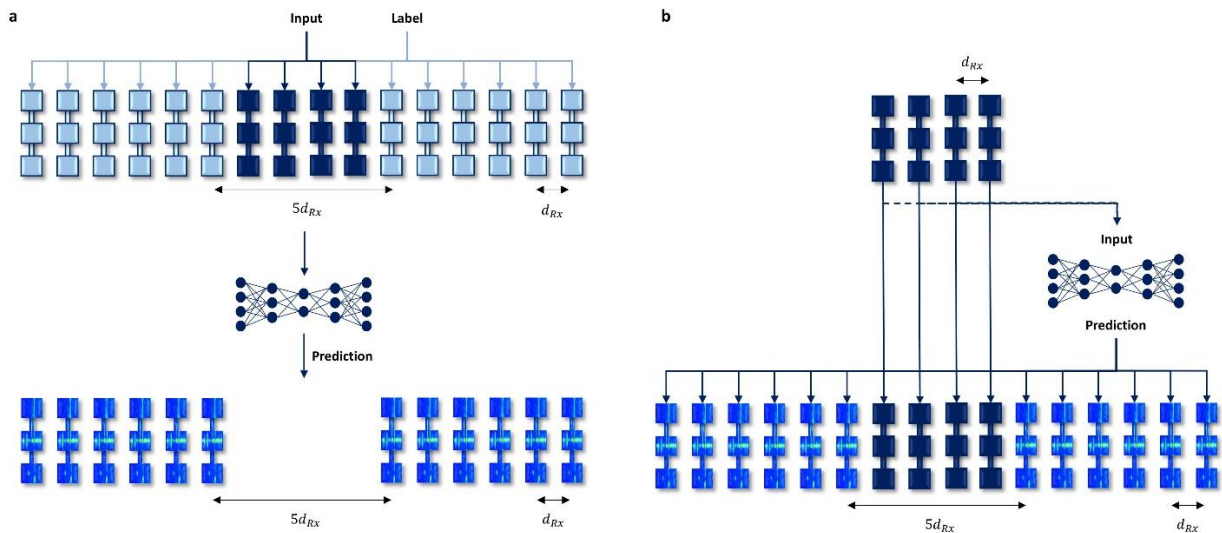

**Fig. 2**. **Radar super-resolution using self-supervised learning a-b. (a)** Training mode: in this example, 4 receiving channels are used as input (dark blue) to predict 12 receiving channels outside of the original aperture (light blue). **(b)** Inference mode: the original array is used as input to a DNN, which predicts adjacent receiving channels outside of the original aperture. Afterwards, both input and predicted receiving channels are used for coherent beamforming. The resulting artificial array has 16 receiving channels and thus 4x improved resolution from the original array.

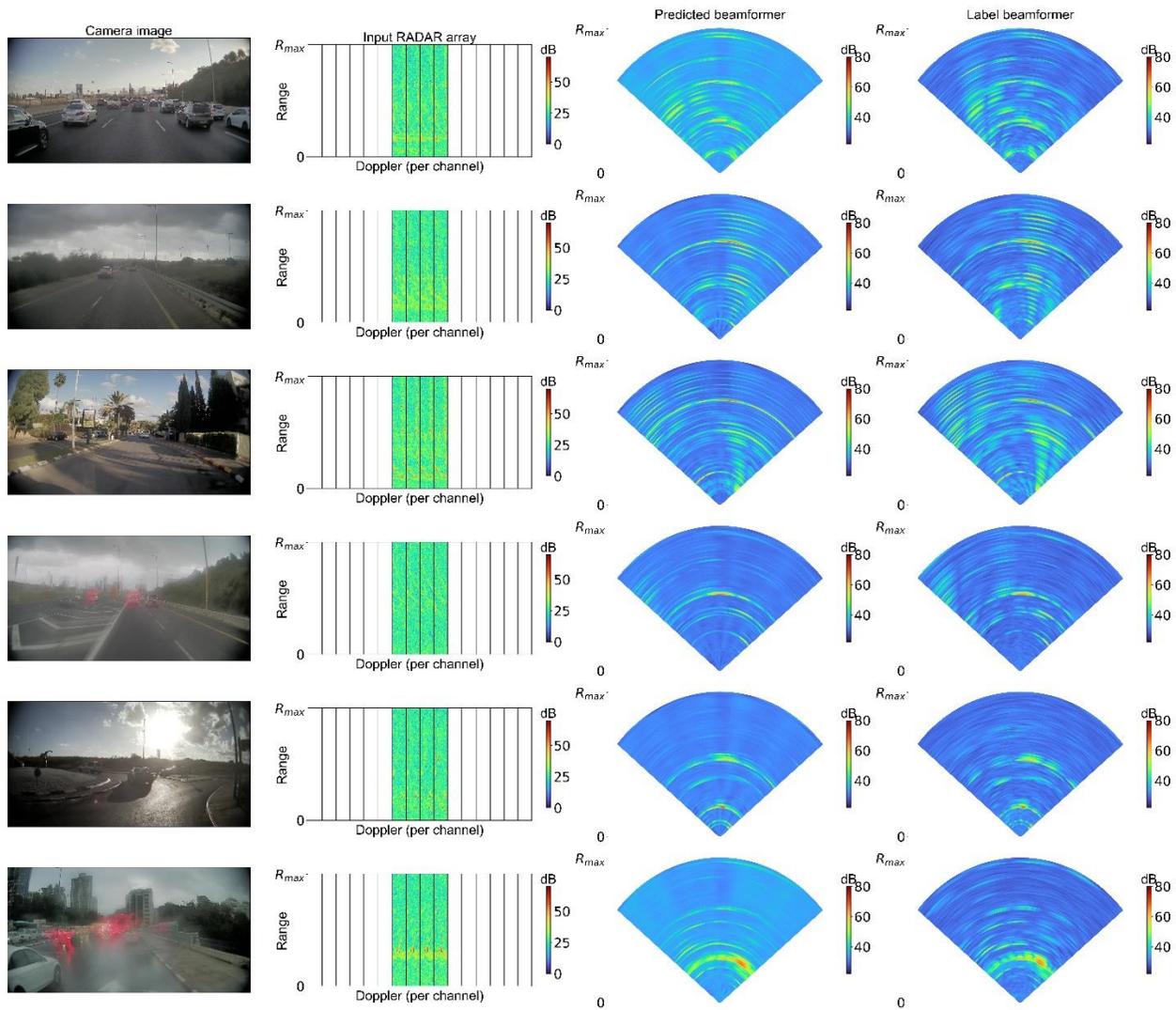

**Fig. 3**. **Sample results from the validation dataset**. From left to right: camera image, input radar array, predicted beamformer and label beamformer. Both beamformers are displayed in cartesian coordinates. The camera image is used for the reader's reference and was not used during training or inference. Blurry images were caused by rain droplets on the camera lens during data collection.

**Fig. 4.** **Detailed results from the validation dataset for 3 representative cases a-c**. Each frame displays a reference camera image, input radar array, predicted radar array and label radar array, with values in dB. Empty spaces were left to orient the reader as to which receiving channel belongs to each group. In addition, range-Doppler Non-Coherent Integration (NCI) is displayed for each array with values in dB and also showing the maximum detection in dotted black lines. The 3 arrays are also displayed in cartesian coordinates with values in dB and a dotted black line signifying the maximum detection range. Three cross sections of the maximum detection are displayed showing the input, predicted and label arrays. In these representative scenarios, the vehicles detections occupy significant angular coverage in the low-resolution radar (input radar array), sometimes blocking an open road, which illustrates the critical need for superior resolution radars. The results also show that by using R2-S2, the input array is super-resolved to match the performance of the label array. (**c**) Displays a sample of a stationary scenario, meaning the radar is not moving, with similar results to samples where the radar was moving. These results suggest the DNN is not relying solely on Doppler and micro-Doppler effects during the reconstruction process. Blurry image was caused by rain droplets on the camera lens during data collection.

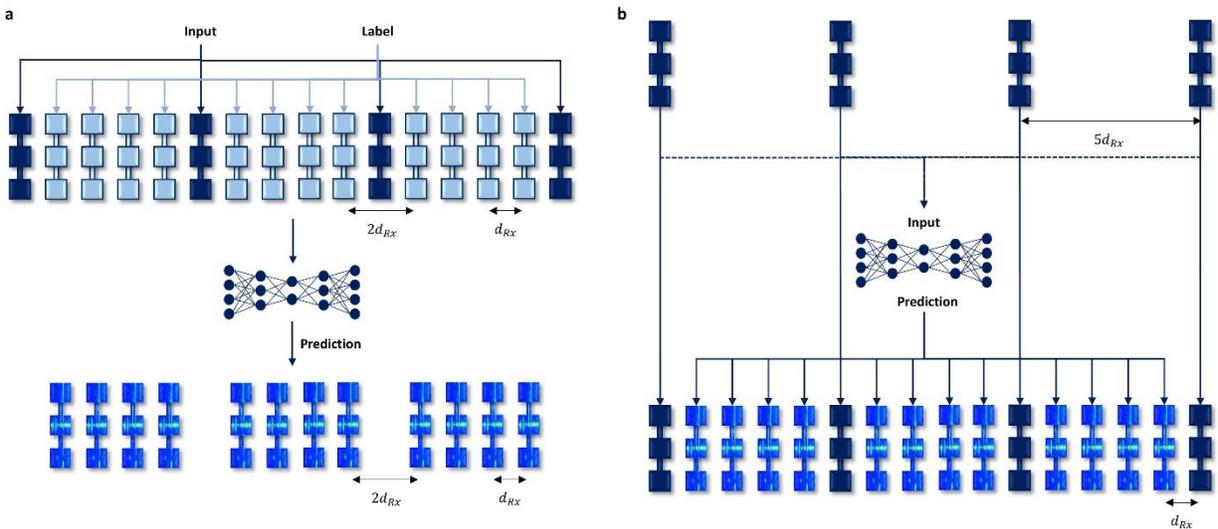

**Fig. 5**. **Coherent beamforming using self-supervised learning a-b**. (**a**) Training mode: in this example, 4 receiving channels (dark blue) are used as input and 12 receiving channels are used as label (light blue). Together they reconstruct a full 16 receiving channel radar array. $d_{Rx}$ is the distance between adjacent receiving channels. (**b**) Inference mode: input receiving channels are first used by a DNN to predict artificial receiving channels, each at specific missing locations in the full array. Afterwards, both input and predicted receiving channels are used for coherent beamforming.

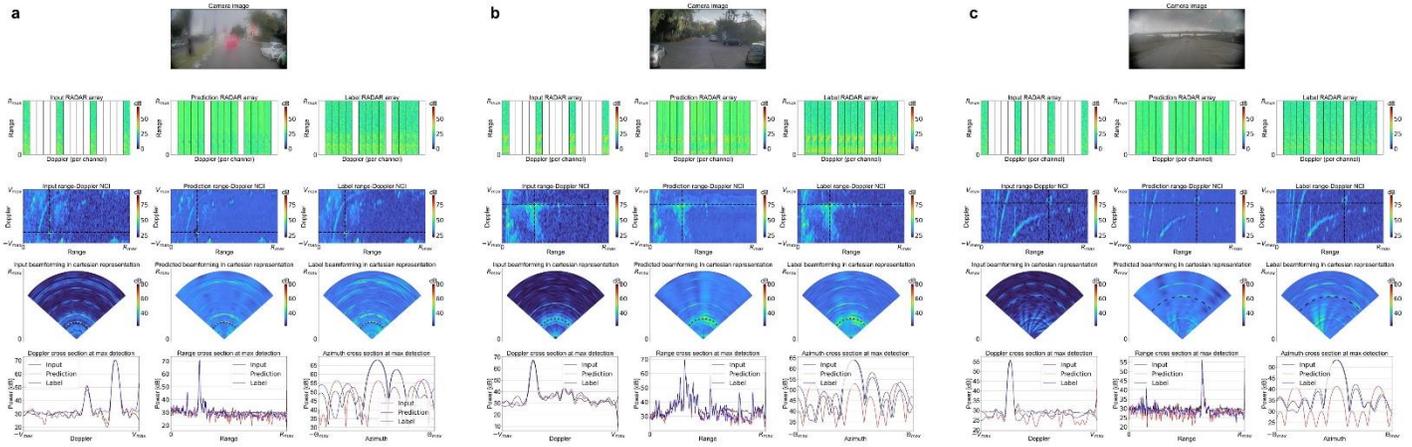

**Fig. 6**. **Sample results from the validation dataset for the Sparse Array configuration a-c**. Each frame displays a reference camera image, input radar array, predicted radar array and label radar array, with values in dB. Empty spaces were left to orient the reader as to which receiving channel belongs to each group. In addition, range-Doppler Non-Coherent Integration (NCI) is displayed for each array with values in dB and also showing the maximum detection in dotted black lines. The 3 arrays are also displayed in cartesian coordinates with values in dB and a dotted black line signifying the maximum detection range. Three cross sections of the maximum detection are displayed showing the input, predicted and label arrays. These results show that beamforming on the input radar array suffers from degraded performance due to grating lobes caused by the large distance between each antenna element. By using R2-S2, the gaps are filled and the performance of the predicted beamformer matches the label beamformer. Blurry images were caused by rain droplets on the camera lens during data collection.

**Tables:**

| | Range-Doppler | | Beamformer | |
|---|---|---|---|---|
| **Loss** | **L1 [a.u.]** | **PSNR [dB]** | **L1 [a.u.]** | **PSNR [dB]** |
| $\mathcal{L}_{rd_{rec}}$ | 0.898 | 30.707 | 0.988 | 42.243 |
| $\mathcal{L}_{rd_{rec}} + \mathcal{L}_{rd_{energy}}$ | 0.868 | 34.673 | 0.959 | 43.324 |
| $\mathcal{L}_{rd}$ | 0.853 | 34.689 | 0.957 | 43.287 |
| $\mathcal{L}_{rd} + \mathcal{L}_{bf_{rec}}$ | 0.811 | 37.789 | 0.843 | 44.766 |
| $\mathcal{L}_{rd} + \mathcal{L}_{bf_{rec}} + \mathcal{L}_{bf_{energy}}$ | 0.806 | 37.886 | 0.846 | 45.148 |
| $\mathcal{L}_{rd} + \mathcal{L}_{bf}$ | **0.796** | **37.991** | **0.794** | **45.584** |

**Table 1. Loss function ablation study**. L1 and PSNR metrics for both range-Doppler representation and beamformer representation (range-Doppler-azimuth). The results were averaged over the validation dataset.

| | Range-Doppler | | Beamformer | |
|---|---|---|---|---|
| | **L1 [a.u.]** | **PSNR [dB]** | **L1 [a.u.]** | **PSNR [dB]** |
| **Bi-cubic** | **0.668** | 38.464 | 1.384 | 48.098 |
| **Our** | 0.734 | **38.649** | **0.886** | **48.956** |

**Table 2. Validation loss metrics for the sparse array configuration**. Lower L1 combined with higher PSNR in the beamformer representation space by our method in comparison to bi-cubic interpolation further suggests coherent reconstruction by R2-S2.

# Supplementary Materials for

## Coherent, super resolved radar beamforming using self-supervised learning

Itai Orr[1, 2, *], Moshik Cohen[2], Harel Damari[2], Meir Halachmi[2], Zalevsky Zalevsky[1]

Correspondence to: itaiorr@gmail.com

**This PDF file includes:**

**Supplementary Text**

<u>Data sparsity examination</u>

An examination on the sparsity of radar data in urban and highway environment was conducted by using the configuration described in Fig 2, where 4 receiving channels are regarded as input radar array and the 16 receiving channels are regarded as the original radar array.

Both radar configurations were compared at each range-Doppler-Azimuth cell for detections above a specific CFAR threshold. Sparse cells were considered as empty or with a single detection. Fig S1 displays data percentage as a function of CFAR threshold for 4 cases: empty cell (red), single detection cell (blue), multiple detections cell (green) and sparse cell (black). Fig. S1b shows static data examination and Fig. S1c shows dynamic data examination.

The results show that the data used in during experimentation especially for static, low CFAR thresholds cannot be consider sparse. The lower sparsity level in the static data is caused due to stationary clutter present in urban scenes.

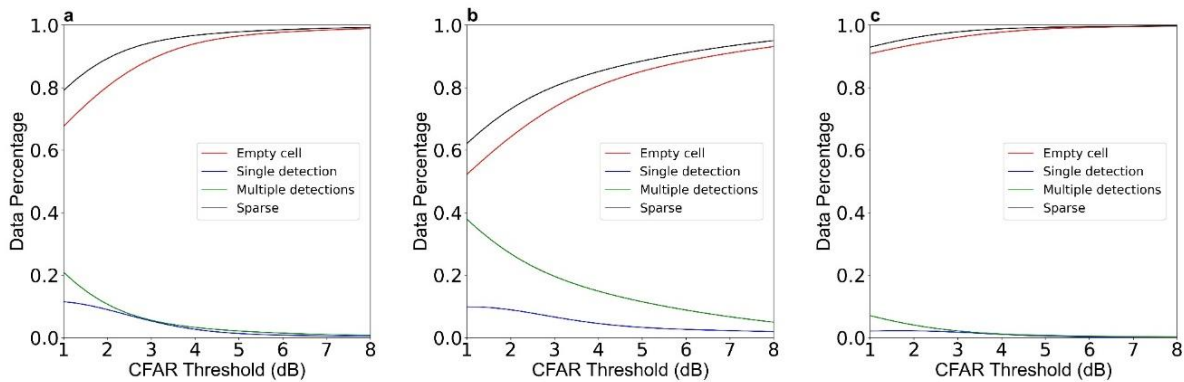

**Fig. S1. Data sparsity examination a-c.** Y axis on a scale of (0-1). **(a)** Data percentage as a function of CFAR threshold for 4 cases: empty cell (red), single detection cell (blue), multiple detections cell (green) and sparse cell (black). **(b)** Static data. **(c)** Dynamic data.

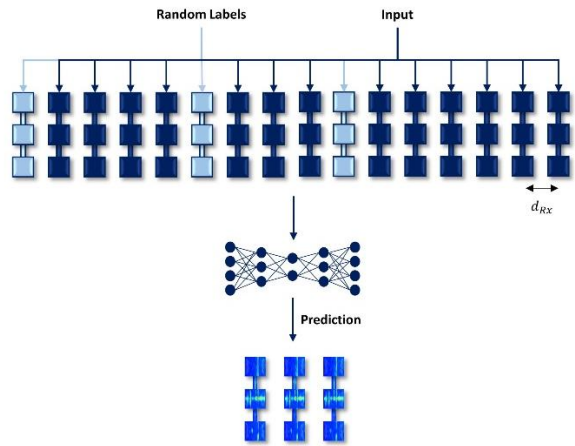

**Fig. S2. Random missing channel configuration.** At training, a random receiving channel is chosen and masked to be used as label. The model then identifies the missing receiving channel and predicts its measurements. Since the missing receiving channel can be located anywhere in the virtual array, this configuration combines both interpolation and extrapolation performed by the model.

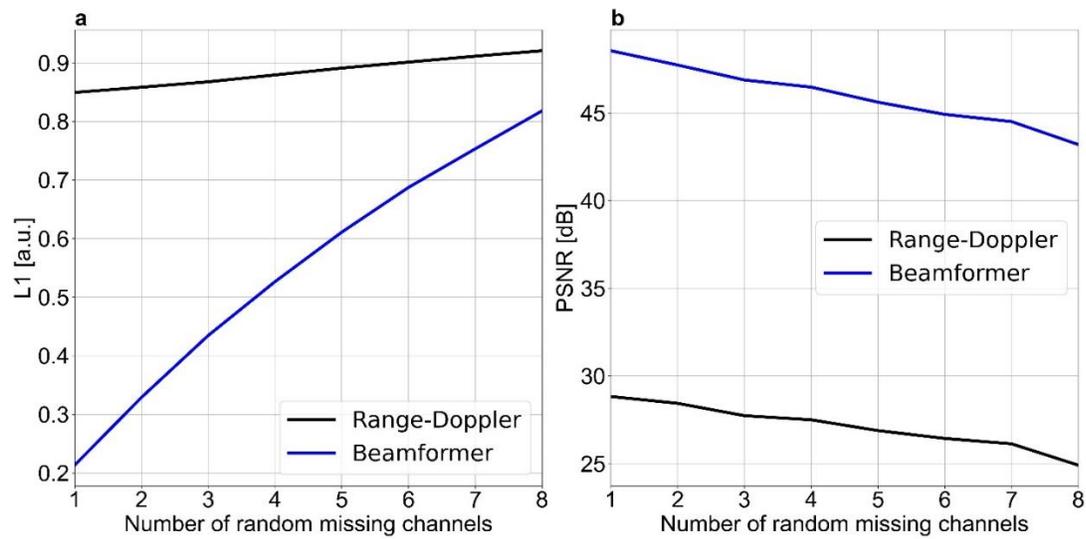

**Fig. S3. Sensitivity analysis of the random missing channels configuration a-b.** We examined the change in L1 loss and PSNR. As expected, there is a performance decrease as the number of random missing channels increase. When combined with a specific performance criterion, this analysis allows to set a maximum number of missing channels the method may be used for in a real-world application.

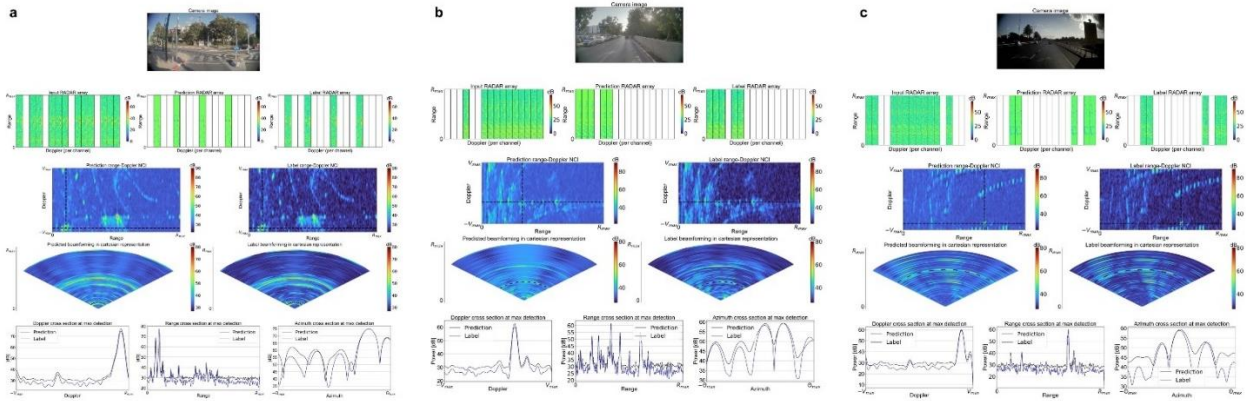

**Fig. S4. Sample results from the validation dataset for the Random Missing Channels configuration a-c.** Each frame displays a reference camera image, input radar array with values in dB and an empty space signifying the missing receiving channel. In addition, range-Doppler maps are displayed for the predicted and label receiving channels with values in dB and also showing the maximum detection in dotted black lines. The predicted and label arrays are also displayed in cartesian coordinates with values in dB and a dotted black line signifying the maximum detection range. Three cross sections of the maximum detection are displayed showing the input, predicted and label arrays. These results show that by using R2-S2 it is possible to overcome randomly missing receiving channels and match the performance of the label array.

**Table S1. Validation Loss metrics for random missing channel configuration**. Note that bi-cubic interpolation cannot be used to estimate the channels at both ends of the array, whereas R2-S2 is able to extrapolate as well as interpolate.

| | Range-Doppler | | Beamformer | |
| --- | --- | --- | --- | --- |
| | L1 [a.u.] | PSNR [dB] | L1 [a.u.] | PSNR [dB] |
| Bi-cubic | 0.879 | 37.360 | 0.251 | 48.412 |
| Our | 0.816 | 31.318 | 0.213 | 48.551 |